\documentstyle[11pt,aaspp4,epsf]{article}

\tightenlines
\newcommand{\r}{\bibitem{}}
\newcommand{\tb}{\mbox{T$_{bol}$}}
\newcommand{\kms}{\mbox{km~s$^{-1}$}}

\begin{document}

\title{\LARGE A Search for Infall Motions Toward Nearby Young
        Stellar Objects}
\author{D.\ Mardones, P.C.\ Myers, M.\ Tafalla, D.J.\ Wilner\altaffilmark{1}}
\affil{Harvard-Smithsonian Center for Astrophysics, 60 Garden St, Cambridge
        MA 02138}
\altaffiltext{1}{Hubble Fellow}
\author{R.\ Bachiller}
\affil{Observatorio Astron\'omico Nacional, Apartado 1143, 
	E-28800 Alcal\'a de Henares, Spain}
\author{G.\ Garay}
\affil{Universidad de Chile, Casilla 36-D, Santiago, Chile}


\begin{abstract}

We report observations of 47 candidate protostars in two optically thick lines
(H$_2$CO $2_{12}-1_{11}$ and CS $2-1$) and one optically thin line (N$_2$H$^+$
$1-0$) using the IRAM 30-m, SEST 15-m, and Haystack 37-m radio telescopes.  The
sources were selected from the redness of their spectra ($\tb < 200$~K) and
their near distance ($d < 400$~pc).  Most of the sources have asymmetric
optically thick lines.  The observed distribution of velocity differences,
$\delta V = (V_{thick}-V_{thin})/\Delta V_{thin}$, is skewed toward negative
(blue-shifted) velocities for both the H$_2$CO and CS samples.  This excess is
much more significant for Class 0 than for Class I sources, suggesting that we
detect infall motions toward Class 0 and not toward Class I sources.  This
indicates a difference in the physical conditions in the circumstellar
envelopes around Class I and Class 0 sources, but does not rule out the
presence of infall onto Class I sources by e.g.\ lower opacity gas.  Bipolar
outflows alone, or rotation alone, cannot reproduce these statistics if the
sample of sources has randomly oriented symmetry axes.  We identify 15
spectroscopic infall candidates, of which 6 are new.  Most of these infall
candidates have primarily turbulent rather than thermal motions, and are
associated with clusters rather than being isolated.

\end{abstract}


\section{Introduction}

Low mass stars form by gravitational collapse of dense cores in molecular
clouds.  The collapse is believed to proceed from a condensed initial state
(Larson 1969, Shu 1977) with a rapid development of a system consisting of a
protostellar core which accretes mass from a disk and envelope.  This
gravitational collapse model can now be explored through detailed observations
and numerical calculations.  Observational studies of star-forming infall are
indispensable to understand the kinematics of gravitational collapse onto young
stellar objects (YSOs).  This, in turn, is essential to study the origin of the
initial mass function and of multiple systems.  Observations are now beginning
to probe a variety of size scales, density and temperature regimes in many
sources.

In recent years, millimeter-wavelength observations have increased the evidence
for infall motions onto nearby YSOs.  The main procedure is to observe spectral
lines that trace high densities ($n > 10^4$~cm$^{-3}$) and have red-shifted
self-absorption spatially concentrated around an embedded YSO. The spatial
concentration and the use of dense gas tracers are important to ensure that the
observed kinematic signature is associated with the YSO.  For instance, Zhou et
al.\ (1993) mapped the globule B335 in the $2_{12}-1_{11}$ and $3_{12}-2_{11}$
optically thick lines of H$_2$CO and in the optically thin lines of C$^{18}$O
$1-0$ and C$^{34}$S $2-1$.  They reproduced the line profiles on spatial scales
of 0.02~pc from the source with an inside-out collapse model (Shu 1977).  Choi
et al.\ (1995) modeled the same B335 observations with a Monte Carlo radiative
transfer code (Bernes 1979), confirming the numerical results from Zhou et al.\
(1993).  Later Zhou (1995) modeled the observed H$_2$CO and CS line profiles
toward IRAS 16293-2422 as arising from a combination of infall and rotation,
using the Terebey, Shu \& Cassen (1984) collapse model.  The case for infall
motions in these sources is strong because of the detailed observations and
modeling that has been done on them.  However, in such a small sample of
candidates, peculiarities of source structure and kinematics (eg, outflows or
rotation) could possibly account for the apparent signatures of infall.
Therefore, these well studied cases are not sufficient by themselves to support
a claim for observed infall motions onto a wide class of sources.

Another approach to search for infall evidence is to observe a statistically
large sample of sources in two lines tracing high density gas, one optically
thick and the other optically thin.  In such a sample, rotation and bipolar
outflow motions having symmetry axes in random orientations would tend to
produce optically thick line profiles with equal numbers of red-shifted and
blue-shifted self-absorption.  On the other hand, infall motions in centrally
concentrated regions would only produce red-shifted self-absorption in the line
profiles (Leung \& Brown 1977).  A spectral line survey of a large source
sample should reveal whether there is a statistically significant excess of
sources with red-shifted or blue-shifted self-absorption.  This approach does
not depend much on the details of the models, but mainly on whether there is a
prevalence of inward or outward motions.  At the same time such a study can
yield a list of collapse candidates for further detailed studies.  Recent
surveys aimed to find candidate collapsing YSOs include Mardones et al.\ (1994)
which used the C$_3$H$_2$ $2_{12}-1_{01}$ line; Wang et al.\ (1995) who
observed the H$_2$CO $3_{12}-2_{11}$ line in a sample of Bok globules, and
Gregersen et al.\ (1997) which used HCO$^+$ 4--3 and 3--2 lines in a sample of
Class 0 sources (Andre, Ward-Thompson \& Barsony 1993).  None of these surveys
has emphasized the statistical properties of the samples because of the limited
number of targets observed.

In this paper we present the results of a molecular line survey toward 47 of
the reddest nearby YSOs, more than twice the number of sources in any of the
previous surveys.  In section 2 we discuss our source selection criteria, the
molecular lines and the telescopes we used.  In section 3 we describe the
normalized velocity difference ($\delta V$) between the peaks of optically
thick and thin lines, as a quantitative measure of the line profiles.  We then
analyze statistically the distribution of $\delta V$ for different molecular
lines and source subsamples.  In section 4 we interpret the statistics
kinematically.  Finally, we summarize our conclusions in section 5.

\section{Observations}

\subsection{Source Sample}

The sources were selected solely based on their observed spectral energy
distribution (SED) and estimated distance from the Sun in order to be unbiased
with respect to prior knowledge of their molecular line shapes.  To select the
youngest sources (best infall candidates) we used the bolometric temperature
(\tb; Myers, \& Ladd 1993) whenever known, otherwise estimated \tb~ based on
IRAS and submm continuum observations.  \tb~ is defined as the temperature of a
black body with the same mean frequency as the source spectrum.  Chen et al.\
(1995) found that Class 0 sources have $\tb < 70$~K, and Class I sources (Lada
\& Wilking 1984) have $70\leq \tb < 650$~K.  We imposed a limit of $\tb <
200$~K to select the most embedded Class I and Class 0 sources.  We relied
mostly on the papers by Ladd, Lada, \& Myers (1993) in Perseus and Chen et al.\
(1995, 1996) in the Taurus, Ophiuchus, Chamaeleon, Lupus and Corona Australis
star forming clouds for determination of \tb.  Chen et al.\ required the source
fluxes to be known in at least 6 different wavelengths in order to compute
accurate values of \tb.  That condition is easily met by most IRAS sources
which have also been detected at optical or near-infrared wavelengths.
However, many of the most embedded Class I sources and most Class 0 sources do
not satisfy that requirement.  Only a handful of Class 0 sources have been
observed in enough different mm and submm wavelengths to make a good
determination of \tb~ possible (eg.\ B335 and L1527, Ladd et al.\ 1991).  We
therefore added sources to our list based on more relaxed criteria in order to
have a larger sample of the youngest known nearby YSOs.

We required IRAS point sources in our sample to have rising spectra between 25
and 100 $\mu$m (with flux ratios 2--4 for wavelength pairs of $25/12$, $60/25$
and $100/60 \mu$m), and to be located close to the centers of their parent
molecular cloud core maps.  These conditions are necessary to exclude galactic
cirrus and extragalactic objects which may have high $100/60 \mu$m flux ratios.
We also added sources with known bright mm or submm continuum emission, and
nearby Class 0 sources known to us by mid 1995.  We consulted Casali et al.\
(1992) for submm continuum observations in Serpens, and Henning et al.\ (1993)
and Reipurth et al.\ (1993) for submm continuum observations toward southern
PMS stars.  We also consulted Persi et al.\ (1990), Carvallo et al.\ (1992) and
Bourke et al.\ (1995) for IRAS and near-infrared observations of southern PMS
stars, many of them located in Bok globules.  We determined \tb~ for all the
sources detected in at least three different wavelengths to include sources not
quoted in Chen et al.\ (1995, 1996).  The derived values of \tb~ for sources
with poor wavelength coverage are upper limits if they lack long wavelength
fluxes, and lower limits if they lack short wavelength fluxes.  Because of our
selection criteria, we are confident that $\tb < 200$~K in all sources in our
sample.  To determine a representative systematic error in \tb~ in sources with
poor short wavelength spectral coverage, we determined \tb~ including and
excluding wavelengths shorter than 60 $\mu$m in sources with good spectral
coverage.  We found that the error in \tb~ when neglecting wavelengths shorter
than 60 $\mu$m is typically $<20$~K for sources having $\tb < 100$~K.  Our
determination of \tb~ is consistent with the values given by Gregersen et al.\
(1997) also with limited wavelength coverage.

Table \ref{Tsources} lists all the sources detected in at least one optically
thin and one optically thick line.  Columns 1 and 2 identify the source,
columns 3 and 4 list the observed coordinates (B1950.0), column 5 lists the
distance from the Sun in pc, column 6 lists \tb, and column 7 the number
N$_\lambda$ of continuum fluxes used to derive \tb.  We give lower or upper
limits to \tb~ rather than values for the sources IRAS 03282+3035, HH211mm,
VLA1623, and S68N due to lack of wavelength coverage close to the peak of their
spectral energy distributions.  Although these estimates are crude, they are
sufficient for our analysis in section 3, since we only need to know whether
\tb~ is below 70~K or not.  Column 8 lists the references we consulted to
determine values of \tb.  Finally, column 9 indicates whether there is outflow
evidence from the source either as high velocity molecular line wings (CO) or
as shock fronts (HH).

   Even though we did not use the outflow properties as a source selection
criterion, 87\% of the sources in Table \ref{Tsources} actually show evidence
of driving an outflow.  This is expected when selecting the most embedded
sources, since they tend to drive strong molecular outflows (Bontemps et al.\
1996).  YSO outflow models usually require mass infall onto an accretion disk
(see references in Bachiller 1996), so there is reason to expect that most
sources in this survey must have infalling envelopes.  However, we emphasize
that in this paper we are seeking to examine direct kinematic evidence for
infall motions onto the sources.

We did not detect 19 of the sources from our initial target list in either the
H$_2$CO line, the N$_2$H$^+$ line, or both.  Table \ref{Tnodet} lists the
positions and rms noise (main beam brightness temperature) of the undetected
lines for reference.  We detected the N$_2$H$^+$ line toward the sources IRAS
11436-6017 and GSS30-1, but it was too weak for a good velocity and width
determination (peak $T_{mb}$ of 0.15 and 0.3~K respectively).  We note that the
average \tb~ of these sources in Table \ref{Tnodet} is higher than that of the
detected sources listed in Table \ref{Tsources}.  In fact, sources with
$\tb <50$~K in Table \ref{Tsources} were easily detected in all the observed
lines.

\subsection{Lines}

It is important to observe both optically thick and optically thin lines in
order to distinguish double-peak profiles produced by optical depth effects
from those due to two velocity components along the line of sight.  The
observed molecules should have a high dipole moment in order to be sensitive to
the high density gas in the immediate neighborhood of the YSO and not to the
more extended, lower density gas surrounding the parent cores.  Also, the
molecular species should be abundant in molecular cloud cores, but unaffected
by peculiar chemistry effects.  For example, SiO and CH$_3$OH are well known
for their enhanced abundances by many orders of magnitude in the outflow lobes
of some YSOs (e.g.\ Bachiller 1996), presumably due to desorption of molecules
from grain mantles in shock fronts.

Following these criteria, we chose the CS ($\mu = 1.96$ Debye) and H$_2$CO
($\mu = 2.33$ Debye) molecules to carry out the survey observations.  H$_2$CO
was our first choice molecule, following the example of Zhou et al.\ (1993).
The abundance of H$_2$CO is known to be only moderately enhanced in bipolar
outflows (Bachiller, \& P\'erez 1997).  We subsequently observed part of the
survey in the CS molecule in order to check that our results were not being
biased by peculiar chemistry in low mass star forming regions.  We believe the
CS and H$_2$CO lines are generally optically thick because of the observed
brightness temperatures and line shapes.  In a few cases, when the line is weak
(T$_{mb}\leq 1$~K) and approximately of Gaussian shape, it is likely to be
optically thin (eg.\ L673A, 04325+2402, L1172; see figs.\ \ref{Fclass0},\ref{FclassI}).

Ideally, one would observe an isotope of the optically thick tracer molecule as
the optically thin tracer.  Then both lines would probe the same density,
temperature and chemistry, and would be observed with essentially the same
angular resolution.  However, the 2-mm H$_2^{13}$CO and the 3-mm C$^{34}$S
lines are in general too weak for accurate determination of line velocities in
a large source sample.  We chose to observe the N$_2$H$^+$ J=$1-0$ line instead
($\mu = 3.4$ Debye).  This molecule has seven hyperfine components, in two
groups of three lines and one isolated component ($JF_1F_2 = 101-012$).
Caselli, Myers, \& Thaddeus (1995) determined the frequencies of all components
with a precision of 6~kHz (0.02 \kms).  The hyperfine structure of N$_2$H$^+$
also permits a determination of the optical depth in the line, thus, allowing
an internal check on the thinness of the isolated N$_2$H$^+$ $101-012$ line.
Caselli et al.\ showed that among the seven components, only two show any
evidence of anomalous excitation ($F_1F_2 = 12-12$ and $F_1F_2 = 10-11$), and
that these anomalies represent a departure of $< 20\%$ from the value expected
in LTE.  Any such small anomaly in hyperfine line ratios is included in the
estimated uncertainty of the optical depth. This uncertainty is generally much
smaller than the estimated optical depths.  Womack, Ziurys, \& Wyckoff (1992)
found that the N$_2$H$^+$ J= $1-0$ line is narrow in dark clouds, with no
evidence of line wings.  The N$_2$H$^+$ line-widths and hyperfine components
make it an ideal probe of the systemic velocity toward young stellar objects.

\subsection{Telescopes}

We used the IRAM 30-m telescope in Pico Veleta, Spain, in 1995, May 10--12.  We
observed simultaneously in the 2 and 3mm atmospheric windows with receivers
tuned to the ortho H$_2$CO $2_{12}-1_{11}$ (140.839518 GHz) and N$_2$H$^+$
$101-012$ (93.176265 GHz) lines respectively.  The line rest frequencies were
taken from Lovas (1991) and Caselli et al (1995) respectively.  The backend had
a digital correlator with spectral resolutions of 20 and 10 kHz at 2 and 3mm
wavelengths respectively (0.04 and 0.03 \kms).  We used overlap frequency
switching by $\pm 3.85$ MHz because this method is efficient and ensures that
the observed line profiles are free of emission from an off-position.  We
checked pointing regularly with continuum sources and found it good to better
than 5$''$.  Typical system temperatures were $\sim$400~K at 2-mm and
$\sim$350~K at 3-mm.  The main-beam efficiency at 93 and 141 GHz was taken to
be 73\% and 55\% respectively.

We used the SEST 15-m telescope in La Silla, Chile, in 1995, September 19--20
and September 24, to observe simultaneously the same 2-mm and 3-mm H$_2$CO and
N$_2$H$^+$ lines as with the IRAM 30-m telescope.  The weather was excellent,
with system temperatures around 150~K in both lines.  We splitted the high
resolution acousto-optical spectrometer into two bands, providing a spectral
resolution of 43 kHz in each band (0.138 \kms at 93 GHz and 0.11 \kms at 141
GHz).  The acousto-optical spectrometer sometimes produces spurious emission
features in the central 3-5 channels; so to avoid interference with the observed
spectral lines, we centered the spectrometer off the rest velocity of each
source by 2--8 km~s$^{-1}$.  We also observed a subsample of sources in the
para H$_2$CO $2_{02}-1_{11}$ transition at 145.602953 GHz and the C$^{34}$S
$2-1$ transition at 96.412982 GHz.  We used main beam efficiencies of 74\% at
93 GHz and 96 GHz, 68\% at 141 GHz, and 67\% at 145 GHz as indicated in the SEST
handbook.

In 1995, December 7--11 and 29--31, we used the NEROC Haystack\footnote{Radio
astronomy observations at the NEROC Haystack Observatory of the Northeast Radio
Observatory Corporation are supported by a grant from the National Science
Foundation.} 37-m telescope to observe the sources visible from Haystack in the
CS 2--1 line at 97.980968 GHz.  We observed left and right circular
polarizations simultaneously with two SIS receivers.  The backend had a digital
correlator which we used with 17.8 MHz bandwidth and 13 kHz spectral resolution
(0.04 km~s$^{-1}$).  Typical system temperatures were 200~K. The antenna main
beam FWHM at 98 GHz (21$''$) compares well with the IRAM 2-mm and 3-mm beams.
We measured a main beam efficiency of $\sim 16\% \pm 3\%$ by scaling spectra of
IRC+10216 to those taken at the IRAM 30-m telescope by Mauersberger et
al.\ (1989).

\section{Results}

\subsection{Variety of Observed Spectral Line Profiles}

\begin{figure}
\epsscale{1.0}
\plotone{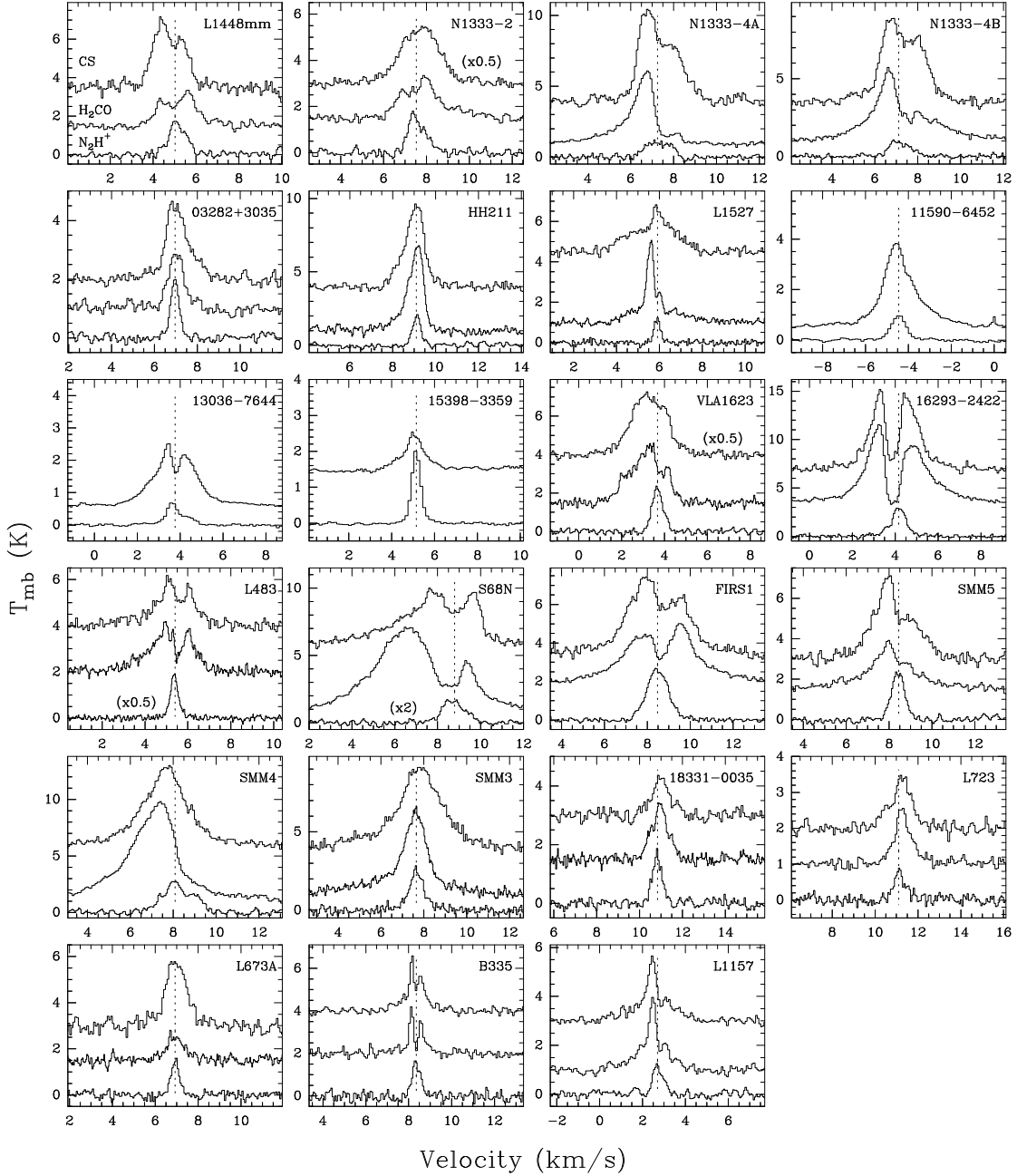}
\caption{Observed spectral line profiles toward the sources with $\tb < 70$~K 
in our sample (``Class 0'').  N$_2$H$^+$ $101-012$ is drawn at the bottom of
each box, H$_2$CO $2_{12}-1_{11}$ next, and CS $2-1$ at the top whenever it
was observed.  The vertical scale is in main beam brightness temperature
units.  The horizontal scale has a range of 10 \kms~ in all sources.  The
dashed line indicates the N$_2$H$^+$ velocity found with the hyperfine
structure fit (the other six hyperfine components lie off the box).
\label{Fclass0}}
\end{figure}

\begin{figure}
\epsscale{1.0}
\plotone{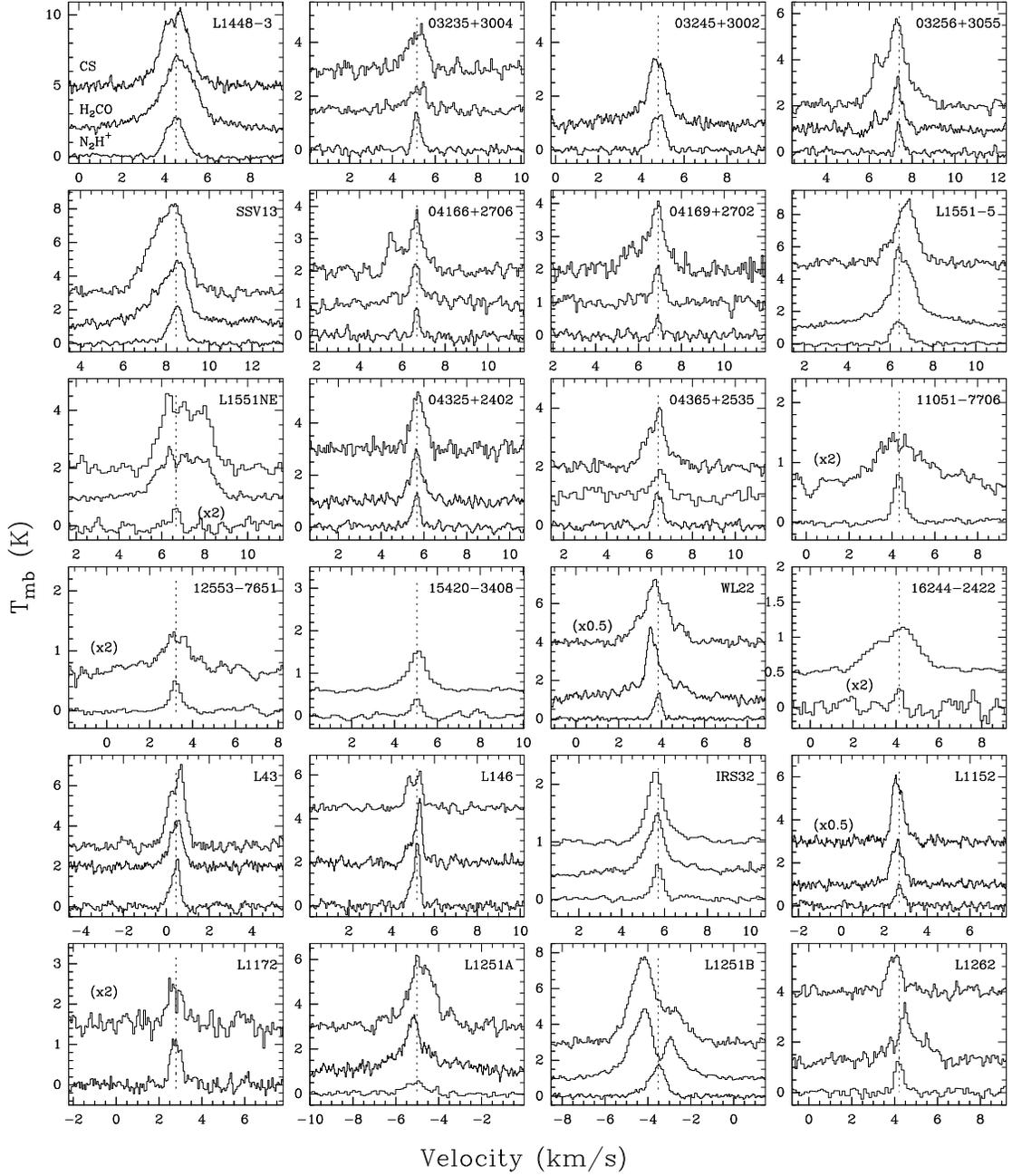}
\caption{Observed spectral line profiles toward the sources with $\tb
\geq 70$~K in our sample (``Class I'').   The display format is the same as in
Fig.\ 1.
\label{FclassI}}
\end{figure}

Figure \ref{Fclass0} shows the observed spectra toward all the Class 0 sources
($\tb < 70$~K) in our sample ordered by right ascension, and figure
\ref{FclassI} the spectra obtained toward the Class I sources ($70 \leq \tb \leq
200$~K).  The N$_2$H$^+$ $101-012$ line is drawn at the bottom of each box;
H$_2$CO $2_{12}-1_{11}$ in the middle, and CS $2-1$ at the top if it was
observed.  In a few sources, such as IRAS 16244-2422, the N$_2$H$^+$ $101-012$
line is very weak, but the velocity (dashed line) is sufficiently well
determined because it is based on all seven hyperfine components, not just the
one shown.  We find a wide variety of spectral line profiles in the H$_2$CO and
CS lines in our sample: double-peaked, with a peak and a ``shoulder,'' and
single symmetric lines.  In addition, many sources have high velocity wings
associated with unbound motions (outflows).

The CS and H$_2$CO line shapes differ from source to source, but are usually
similar to each other.  The N$_2$H$^+$ line, on the other hand, is Gaussian
toward almost all sources.  In sources with symmetric H$_2$CO and CS lines,
their peak velocity lies very close to that of the N$_2$H$^+$ line.  In sources
with double peaked H$_2$CO and CS lines, the N$_2$H$^+$ peak velocity lies
between the two peaks (or between the peak and the shoulder), indicating that
the complex H$_2$CO and CS line profiles arise from self-absorption at low
velocities.  As a proof of the lower optical depth of N$_2$H$^+$ we derive
optical depths $\tau < 1$ in the $101-012$ line in 93\% of the sources, and
$\tau < 1.6$ in all of them.

In most Class I sources, the H$_2$CO and CS lines are broader than the
N$_2$H$^+$ line, but all three line profiles are essentially symmetric showing
no evidence of self-absorption (Fig.\ \ref{FclassI}).  On the other hand, many
of the observed CS and H$_2$CO profiles toward the Class 0 sources show
evidence of self-absorption.  The H$_2$CO and/or CS spectral profiles are
asymmetric with respect to the N$_2$H$^+$ center velocity in 15 out of 23 Class
0 sources (Fig.\ \ref{Fclass0}).  Most of these asymmetric profiles have peaks
whose velocity is significantly blue-shifted, rather than red-shifted, from the
velocity of the optically thin N$_2$H$^+$ line.  This high incidence of
self-absorption toward Class 0 sources is similar to what Gregersen et al.\
(1997) found in HCO$^+$ lines.

Besides the variation in line shapes discussed above, the line widths vary by a
factor of five over our source sample.  For example, the FWHM of the
N$_{2}$H$^{+}$ line varies from 0.3 \kms~ in Taurus to 1.8 \kms~ in Serpens.
This is important to bear in mind when we use the peak velocity shift between
lines to compare different sources in the following section.

\subsection{Quantifying Spectral Line Asymmetries}

To study the distribution of line profiles statistically, we need to quantify
the observed line asymmetries.  To do this, we define the non-dimensional
parameter
\begin{equation}
\delta V = (V_{thick}-V_{thin}) /\Delta V_{thin}, 
\label{dveq}
\end{equation}
the velocity difference between the peaks of the optically thick and thin
lines, normalized by the FWHM of the thin line.  We use the normalized velocity
difference $\delta V$ rather than the velocity difference $V_{thin}-V_{thick}$
to reduce bias arising from lines having different width from one source to the
next.  This is a significant effect in our sample, as discussed above, and the
normalization ensures that two lines of the same shape will have the same
measure of asymmetry even if they have significantly different widths.  To
compute $\delta V$, we first fit a Gaussian to the brightest peak of the
optically thick line to define its velocity ($V_{thick}$).  We then fit seven
Gaussian hyperfine components to the N$_2$H$^+$ $J=1-0$ line using the line
parameters derived by Caselli et al.\ (1995); this gives the N$_2$H$^+$
velocity and FWHM ($V_{thin}$ and $\Delta V_{thin}$).

We believe that $\delta V$ is a more convenient and robust measure of the line
asymmetry than the even-odd component decomposition (Mardones et al.\ 1994),
the shift in line centroid (Adelson, \& Leung 1988; Walker, Narayanan, \& Boss
1994), the ratio of blue to red peak brightness (Myers et al.\ 1995), and the
line skewness (Gregersen et al.\ 1997).  The even-odd decomposition is too
sensitive to the choice of reference velocity, and the line centroid is too
sensitive to the choice of velocity window.  The ratio of blue to red peak
brightness cannot be used for lines where a second peak is not well defined.
The skewness is much more sensitive than is $\delta V$ to asymmetry in the line
wings, which may be due to outflow, not infall motions.  On the other hand, the
adopted $\delta V$ has the disadvantage that it is ambiguous for double-peaked
lines with nearly equal peak intensities.  In our sample this occurs in only 3
sources in the CS line (IRAS 16293-2422, L483 and Serp-S68N) and one source in
the H$_2$CO line (L483; see fig.\ \ref{Fclass0}).  In the CS spectra, the ratio
of the blue--red temperature difference to the rms noise ($(T_b-T_r)/\sigma$)
is 3.7, 3.0, and 0.8 in the sources IRAS 16293-2422, L483, and Serp-S68N
respectively, and in the L483 H$_2$CO spectrum the same ratio is 2.4.   We
regard a ratio of 2.0 or less as insignificant, so we drop the CS line in
Serp-S68N from the sample.

We investigated alternatives to fitting Gaussians to obtain the velocity of
peak emission for the optically thick H$_2$CO and CS lines.  We explored
measuring the velocity of the peaks directly from the spectra, calculating the
moments of the spectra, and finding the velocity of the peak of the
cross-correlation of the optically thick and thin line profiles.  For simple
lines all methods agree to within the rms uncertainty in center velocity due
to noise.  We adopt the direct fit method because it appears slightly more
robust than the cross-correlation or first moment methods for sources with
significant wing emission.  In the optically thin N$_2$H$^+$ J=1-0 line, the
results of fitting seven Gaussian hyperfine components and of fitting a single
Gaussian to the isolated $101-012$ line are consistent to within 0.01 \kms,
but the error in the former is smaller.  

Table \ref{Tlines} lists our main survey data: the observed velocities in the
ortho-H$_2$CO $2_{12}-1_{11}$, CS $2-1$, and N$_2$H$^+$ $1-0$ lines, and the
values of $\delta V_{H_2CO} = (V_{H_2CO}-V_{N_2H^+}) /\Delta V_{N_2H^+}$ and
$\delta V_{CS} = (V_{CS}-V_{N_2H^+}) /\Delta V_{N_2H^+}$ computed from these
velocities, as in equation \ref{dveq}.  Table \ref{Tisotope} lists the observed
velocities in the smaller sample of para-H$_2$CO $2_{02}-1_{01}$ and C$^{34}$S
$2-1$ lines, which we observed as consistency checks.  The velocity error
estimates from the fitting routines are usually smaller than the channel width
(except in low signal to noise spectra).  We quote either one half the channel
width or the Gaussian estimate as the error in our peak velocity
determinations, whichever is bigger.  The error in $V_{thin}-V_{thick}$ is
usually dominated by the error in the optically thick line profile velocity.
This error is typically $\sim$0.03 km~s$^{-1}$ and is always smaller than 0.11
km~s$^{-1}$.

\subsection{Distributions of Normalized Velocity Differences \label{Sdvdist}}

\begin{figure}
\epsscale{1.0}
\plotone{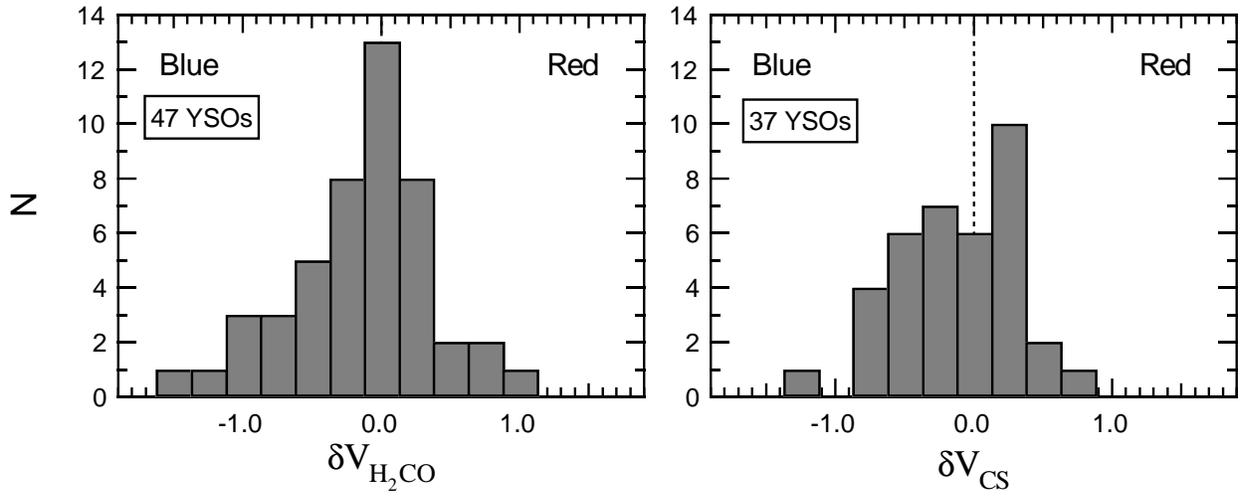}
\caption{Distribution of the line peak velocity difference $\delta V$, defined
in eq.\ \ref{dveq}, for the full sample using: H$_{2}$CO ({\em left}), and CS
({\em right}) as the optically thick line tracer.  Note that more sources have
negative (blue) than positive (red) values of $\delta V$.
\label{Fhisto}}
\end{figure}

Figure \ref{Fhisto} shows histograms of the distribution of $\delta V_{H_2CO}$
and $\delta V_{CS}$.  The IRAM observations were used in figure \ref{Fhisto}
whenever sources were observed with both the SEST and IRAM telescopes because
of the better spectral and spatial resolution (ie.\ each source is counted
once).  The histograms are clearly skewed towards negative velocities.  This
can also be seen from the means and standard errors of the mean of the
distributions $\langle \delta V_{H_2CO} \rangle = -0.15 \pm 0.08$, and $\langle
\delta V_{CS} \rangle = -0.16 \pm 0.07$, given in Table \ref{Tstat}.  We now
explore the statistical significance of this result.

We use the Student's t-test for a sample with unknown spread to compare the
$\delta V$ distributions with a zero-mean normal distribution (our null
hypothesis), and derive the probability (p) of the null hypothesis. The
probability of drawing the distribution of $\delta V$ from a zero mean normal
parent distribution is only 5\% (3\%) for the H$_2$CO (CS) sample.

Another way to quantify the asymmetries in the distribution of $\delta V$ is to
simply count the excess (E) number of sources with negative compared to
positive $\delta V$.  We define the ``blue excess'' E as $(N_{-}-N_{+})/N$,
where $N_{-}$ is the number of sources with a ``significant'' blue-shift
($\delta V < -0.25$), $N_{+}$ is the number of sources with a ``significant''
red-shift ($\delta V > 0.25$), $N_0$ is the number of sources with no
significant shift ($-0.25 \leq \delta V \leq 0.25$), and $N$ is the total
number of sources.  Defined in this way, a large excess (E) corresponds to a
small probability (p).  We chose the threshold value 0.25 in the definition of
E at about five times the typical rms error in $\delta V$ (Table \ref{Tlines})
to screen out random contributions.  The excess of sources with negative
$\delta V$ in the histograms in figure \ref{Fhisto} is 21\% and 29\% for the
H$_2$CO and CS samples respectively (25\% overall).

To understand the origin of the asymmetry of the $\delta V$ distribution, we
evaluate statistics for subsamples defined by whether \tb~ corresponds to a
Class 0 or a Class I source (from here on, we refer to sources with $\tb <
70$~K as Class 0, and to sources with $70 \leq \tb \leq 200$~K as Class I).
Table \ref{Tstat} lists (for the full sample and these subsamples) the
quantities $N$; $N_{-}$,$N_{0}$,$N_{+}$; the mean; p, and E for the
distributions of $\delta V_{H_2CO}$ and $\delta V_{CS}$.

\begin{figure}
\epsscale{0.9}
\plotone{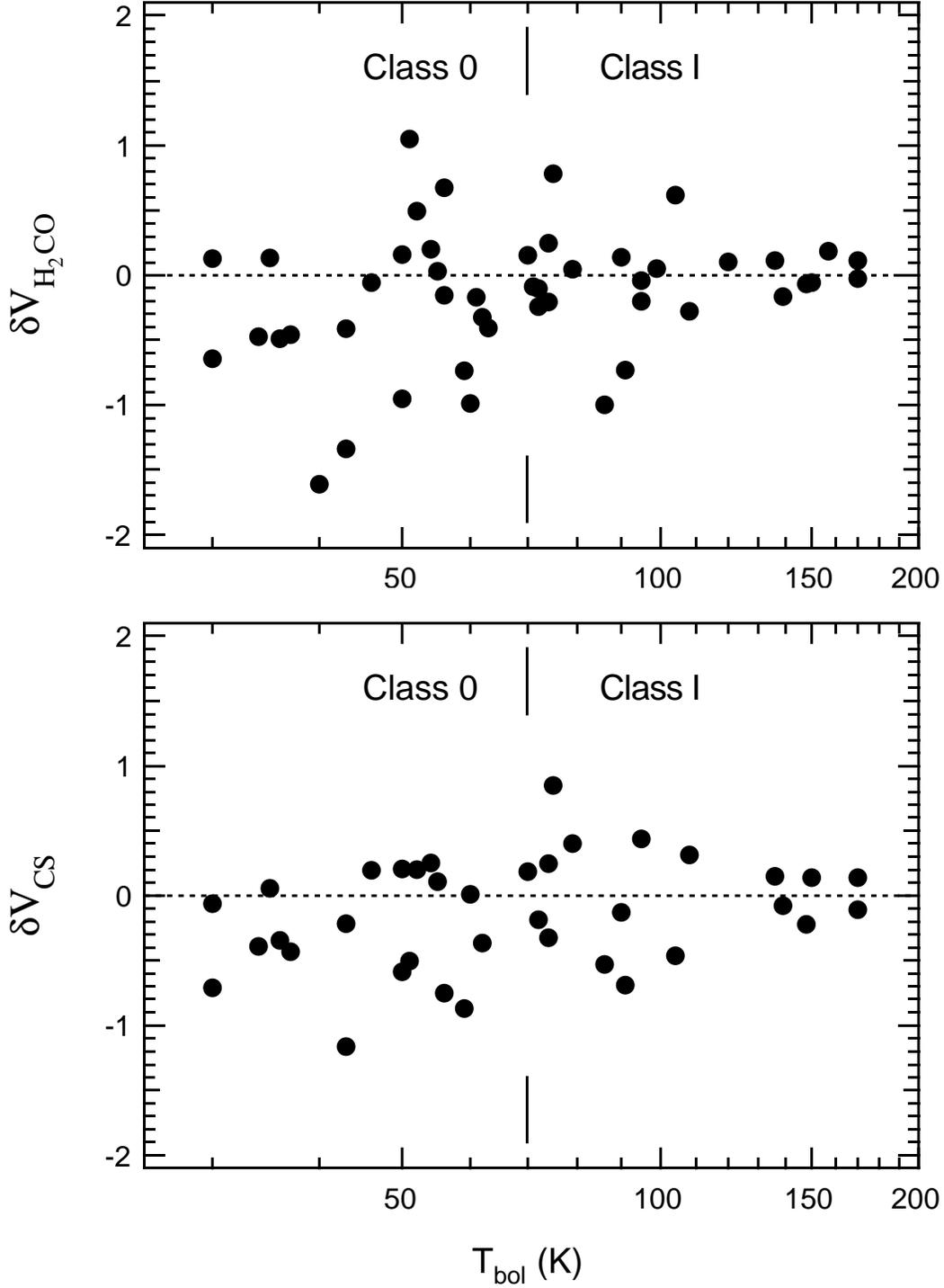}
\caption{Relation between the velocity difference $\delta V$ based on
H$_2$CO ({\em top})and CS ({\em bottom}) as the optically thick line, and
bolometric temperature (\tb).  Both plots show that most sources have
blue-shifted values of $\delta V$ at low \tb~ (Class 0 sources).  At higher
\tb~ (Class I sources) there are equal number of blue- and red-shifted sources.
The uncertainty in $\delta V$ is typically $\sim$0.05 \kms~ (Table
\ref{Tlines}), and the uncertainty in \tb~ is typically of order 20~K (see
text).
\label{Fdvtb}}
\end{figure}

Table \ref{Tstat} shows that the sample sources with $\tb < 70$~K is {\it
entirely } responsible for the histogram skewness in figure \ref{Fhisto}.
Thus, Class 0 sources have a probability of being drawn from a zero-mean normal
distribution of 4\%(0.4\%) for the H$_2$CO (CS) sample, while Class I sources
have a probability of 81\% (99\%) for the H$_2$CO (CS) sample.  The overall
excess of 25\% of sources with negative $\delta V$ can be decomposed in a 47\%
excess (39\% for H$_2$CO and 55\% for CS) among the Class 0 sources, and 2\%
excess (4\% and 0\% respectively) among the Class I sources.  This is
illustrated in Fig.\ \ref{Fdvtb}, which shows the $\delta V$ distributions as a
function of \tb~ for the H$_{2}$CO and CS lines.  The fraction of blue-shifted
sources (negative $\delta V$) in figure \ref{Fdvtb} is clearly greater for
class 0 than for Class I sources.

%

To summarize, sources with $\tb < 200$~K (Class 0 and I) tend to have
asymmetric profiles in both of the optically thick lines that we used to trace
high densities ($>10^4$ cm$^{-3}$).  In this sample there are $\sim$25\% more
sources with a significant blue-shift than with a significant red-shift.  This
statistically significant tendency arises entirely from sources with $\tb <
70$~K.  In this Class 0 subsample the blue-shifted excess is $\sim$50\%, while
in the Class I subsample there is no significant blue- or red-shifted excess.

\subsection{Uncertainties Due to Choice of Line and to Angular Resolution}

Our statistical results are significant as long as they are independent 
of the instruments used, or the lines observed. We compare among different
lines to better understand the sources of uncertainty in the statistical
results discussed above, in particular possible effects of chemistry and
resolution.  We compare the optically thick line profiles of ortho and para
H$_2$CO, and CS; and the optically thin line profiles of N$_2$H$^+$ and
C$^{34}$S.

\subsubsection{Molecular Tracers as a Source of Uncertainty}

\begin{figure}
\epsscale{0.9}
\plotone{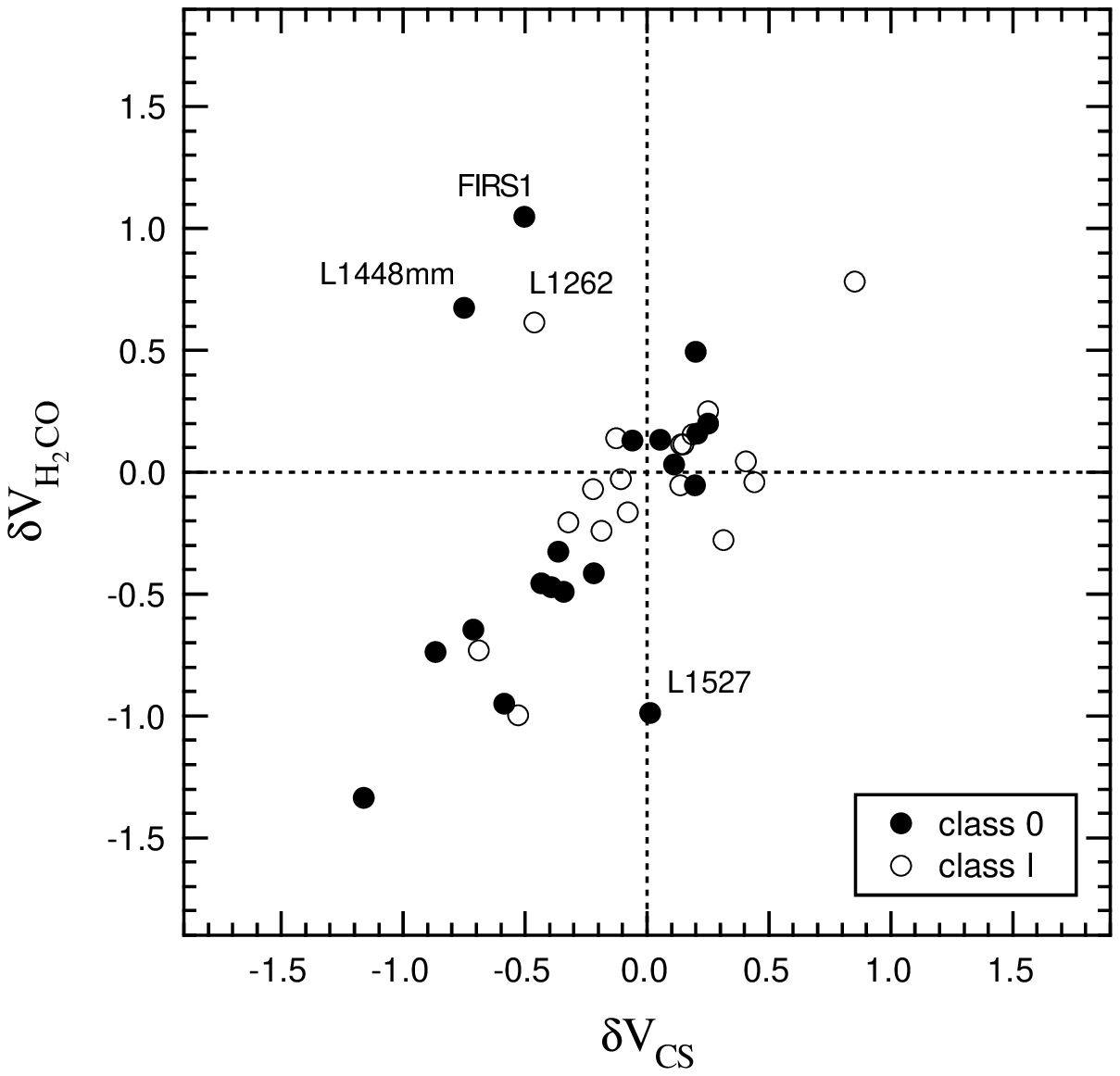}
\caption{Relation between $\delta V$ determined using CS $2-1$, and $\delta V$
determined using H$_2$CO $2_{12}-1_{11}$ as the optically thick lines; and
using N$_2$H$^+$ $101-012$ in each case as the optically thin line.  Note the
good correlation in most sources.  The uncertainty in $\delta V$ is typically
$\sim$0.05 \kms~ (Table \ref{Tlines}), comparable to the symbol size.
\label{Fh2cocs}}
\end{figure}

The H$_2$CO and CS observations give statistically similar results as shown in
section \ref{Sdvdist}; we now compare the line profiles individually in each
source.  The measured peak velocities in the CS $2-1$ and H$_2$CO
$2_{12}-1_{11}$ lines agree quite well as can be seen by inspection of Table
\ref{Tlines}.  Figure \ref{Fh2cocs} plots $\delta V_{H_2CO}$ against $\delta
V_{CS}$ .  Of the 39 sources observed in both lines, 4 (L1448mm, Serp-FIRS1,
L1262 and L1527) deviate significantly from the line of perfect correlation.
For the remaining 35, the best fit line has a slope of $1.05 \pm 0.09$, an
intercept of $-0.09 \pm 0.04$ and a correlation coefficient of 0.90.  The
small differences in the statistics derived above (see Table \ref{Tstat}) arise
mostly from the different CS and H$_2$CO lines toward the four discrepant
sources mentioned above.  A paired t-test yields a probability of 0.91 that
the means of the CS and H$_2$CO $\delta V$ distributions are equal, so we
treat both distributions as being drawn from the same parent distribution.  We
note that the CS and H$_2$CO velocity differences were normalized by the same
N$_2$H$^+$ line width, so the correlation in figure \ref{Fh2cocs} is
independent of the normalization.


We observed the C$^{34}$S $2-1$ line toward 11 sources with the SEST and 2
sources with the IRAM telescope (table \ref{Tisotope}).  We evaluate the effect
of using N$_2$H$^+$ instead of C$^{34}$S to compute $\delta V$ with CS $2-1$ as
the thick line.  The difference $\delta V_{C^{34}S} - \delta V_{N_2H^+}$ has
mean $\pm$ standard error of the mean (s.e.m.) of $0.06 \pm 0.05$, smaller
than the width of one bin in figure \ref{Fhisto}.  Most of the variation in
$\delta V$ comes from the different widths of C$^{34}$S and N$_2$H$^+$ in some
sources.  The C$^{34}$S lines are broader than the N$_2$H$^+$ lines by a factor
of $1.3 \pm 0.1$.  The velocity difference has mean $\pm$ s.e.m\ $0.02 \pm
0.03$ \kms~ ($V_{C^{34}S} - V_{N_2H^+}$).  Only Serp-FIRS1 and L1251B
have C$^{34}$S velocities differing by more than 0.1 km~s$^{-1}$ from the
N$_2$H$^+$ velocity.


Ortho- and para-H$_2$CO are expected to trace the same physical conditions, and
in LTE they are expected to have an ortho/para abundance ratio equal to three.
The observed ortho and para line profiles are very similar in most sources,
indicating high optical depth in both lines.  In particular, the velocity of
the peak is essentially the same in both transitions.  The change in $\delta V$
for all 12 sources has mean $\pm$ s.e.m.\ $0.09 \pm 0.05$ (para -- ortho),
smaller than the width of one bin in figure \ref{Fhisto}.  The ortho- and
para-H$_2$CO lines have different shapes only toward the sources L483 and
VLA1623, where the ortho-H$_2$CO line peak velocity is more blue-shifted with
respect to the N$_2$H$^+$ velocity than is the para-H$_2$CO line.

\subsubsection{Angular Resolution as a Source of Uncertainty \label{Sres}}


We explored the effects of different beam sizes by comparing the IRAM H$_2$CO
line profiles with average profiles over an effective 37$''$ beam ($3\times 3$
points with 10$''$ spacing and 16$''$ beam) in a sample of 13 sources where we
have made maps (Mardones et al.\ 1997).  The self-absorption depth and blue to
red peak ratio vary somewhat, but the peak velocities are consistent to within
0.1 \kms.  The only exception is Serpens-FIRS1, where the red peak is brighter
in the IRAM central spectrum but the blue peak is brighter in the IRAM $3\times
3$ average.  The difference in $\delta V$ between the average spectrum of the
$3\times 3$ map and the center spectrum has mean $\pm$ s.e.m.\ $0.04 \pm 0.03$.
In general, then, the measured $\delta V$ is not very sensitive to variations
in pointing or spatial resolution.


We also compared SEST and IRAM observations to study the variation of the line
profiles when changing spectral and spatial resolution.  The observed SEST line
profiles are on average 40\% fainter than the corresponding IRAM line profile.
This is likely due to beam dilution indicating that the emitting gas is
spatially concentrated on scales smaller than the SEST main beam (36$''$).
Some of the sources proved to be very compact as judged by the high IRAM/SEST
intensity ratio (notably 16293-2422 and B335).  The shapes of the lines
observed with both telescopes are very similar in most of the sources, with a
few exceptions noted below.  For sources with double-peaked H$_2$CO line
profiles the overall intensity and width of the lines is similar at both
telescopes, but the self-absorption depth and the ratio of the two peaks can
vary significantly.  The dips are always deeper in the IRAM spectra, maybe
indicating that the absorption is also diluted in the SEST beam.  The peak
velocities measured in the SEST and IRAM spectra are consistent with each
other.  The change in $\delta V$ using the H$_2$CO velocities measured at the
SEST and IRAM has mean $\pm$ s.e.m.\ $0.01 \pm 0.05$ (SEST-IRAM), for the 10
sources in common, excluding Serp-FIRS1.  The SEST spectrum toward Serp-FIRS1
is almost identical to the IRAM $3'\times 3'$ average spectrum.  We find no
telescope biases in our observations that could affect significantly the
statistics.

\section{Discussion}

We have found an excess of sources with asymmetric optically thick lines
(double-peaked or shoulder profiles) toward a sample of 47 nearby low mass
YSOs.  This asymmetry is such that the peak velocity of the optically thick
line is preferentially blue-shifted with respect to the optically thin
velocity.  This blue-shifted excess is statistically more significant in
subsamples of sources with $\tb < 70$~K.  The observed blue excess of $\delta
V$ values is similar for the CS and H$_2$CO tracers and is independent of the
telescope used.  It is, therefore, an inherent property of the sources in the
samples.

\subsection{Kinematic Interpretation of the Observed Velocity Differences}

The observed N$_2$H$^+$ line profile is single in all sources observed, ruling 
out multiple components as the cause of the asymmetric lines observed.  In a 
few cases (eg.\ L1448mm), the N$_2$H$^+$ line profile departs significantly 
from a Gaussian.  However, in those cases the peak of the N$_2$H$^+$ line 
profile lies within the H$_2$CO or CS absorption dip and does not coincide
with the peak of the H$_2$CO or CS profile.  


A foreground absorbing layer could reproduce the asymmetric line profiles we
observe.  However, we do not expect such a layer of dense gas (having the
systemic velocity of the source) to lie in front of most of the sources
observed unless it is physically associated.  On the other hand, the spatial
variation of optically thick line profiles show that the asymmetries are
concentrated toward the sources (\S \ref{Sres}; also Zhou et al.\ 1993, Myers
et al.\ 1995, Mardones et al.\ 1997) making the casual superposition of two
clouds even more unlikely.  We are led to explain the variety of line profiles
observed as arising from {\it local } self-absorption, i.e.\ absorption by
physically associated gas.

We now ask what kinematics should the local absorbing gas have with respect to
the YSO to reproduce the observed $\delta V$ distributions.  If the kinematics
around a YSO are dominated by infall, then we expect the foreground gas to be
preferentially red-shifted with respect to the systemic velocity of the
source.  If this foreground gas has colder excitation temperature than the
dense gas in the inner cloud, it will absorb preferentially at red-shifted
velocities causing asymmetric line profiles (eg Leung \& Brown 1977; Zhou
1992).  Thus it will have a distribution of $\delta V$ skewed toward blue
velocities as observed.  If systematic motions are due solely to infall toward
a high fraction of Class 0 sources, these motions can explain the observed
statistics.

The observed blue excess of $\delta V$ can arise from infall, as opposed to
expansion; but other motions, including rotation and bipolar outflows, need to
be considered.  Adelson \& Leung (1988) showed that rotation can cause
infall-like and expansion-like line profiles on either side of the rotation
axis (see also Zhou 1995).  The observed effect depends on the orientation of
the rotation axis with respect to the line of sight.  A source sample with
randomly oriented rotation axes would produce a symmetric distribution of
$\delta V$.  Similarly, bipolar outflows also cause infall- and expansion-like
line profiles on either side of the source (Cabrit, \& Bertout 1986).  An
ensemble of bipolar outflows with random orientation would also produce a
symmetric distribution of $\delta V$, unlike our results.  Thus our
statistical results do not rule out the contribution of outflow and rotation
to the line profiles, but they require inward motions in addition to any
outflow and rotation in a significant fraction of our source sample.

Outflow motions greatly affect the observed line profiles toward some sources,
in particular among the Class 0 sample.  If outflow axes in this sample were
not distributed randomly, or if the sample were too small, the outflows could,
in principle, produce the observed statistical effect.  We now examine on a
source-by-source basis whether the observed outflow wings are associated with
the excess of sources with blue $\delta V$ in our sample.  We subtracted one
or two best-fit Gaussians from the H$_2$CO and CS line peaks to measure the
remaining blue and red wings in every spectrum.  We then labeled each source's
wings as blue if the blue wing is $>$25\% brighter than the red, red if the
red wing is $>$25\% brighter than the blue, and nil otherwise.  We find as
many sources with brighter blue and red wings ($\sim 25\%$ of each) in both
the H$_2$CO and CS lines, therefore the excess of sources with negative
$\delta V$ {\sl is not} associated with a similar excess of sources with
bright blue wings.

To test this further, we defined a wing parameter with values -1, 0, or 1 for
sources with brighter wings to the blue, neither, or red respectively, and a
line-core parameter with values -1, 0, or 1 for sources with a measured
$\delta V< -0.25$, $-0.25 <\delta V <0.25$, or $\delta V >0.25$ respectively.
We find a Spearman rank correlation coefficient between the wing and the
line-core parameters of 0.0 for the H$_2$CO sample and -0.1 for the CS sample.
We conclude that the observed wing emission is not correlated with, and thus
is not responsible for the excess of sources with negative $\delta V$ derived
in section 3.

Finally, it is possible that star forming groups could have large scale
rotation or magnetic fields imposing a characteristic orientation to the disks
and outflows around YSOs.  This effect could bias the statistics within each
cluster.  This is evidently not the case in this survey, because the projected
outflow axes have a wide distribution in position angle within each cluster.
The distribution of axis position angles of all outflows with well known
orientation in NGC1333 has an rms of 56\arcdeg (Bally, Devine, \& Reipurth
1996).  Bontemps et al.\ (1996) found a similar range of outflow axis position
angles, with an rms of 55\arcdeg in Ophiuchus.  In Serpens, the position angles
of the outflows axis of the sources S68N, FIRS1, SMM2, SMM4, and SMM5 in
Serpens have an rms of 43\arcdeg (White, Casali,, \& Eiroa 1995, Williams et
al.\ 1997).  Therefore, within each of the clusters in our survey we can regard
the source outflow axes as randomly oriented.


\subsection{Infall Toward Class 0 and Class I Sources \label{S0I}}

We have shown that the observed distribution of $\delta V$ is explained best by
motions which include infall in addition to outflow and/or rotation.
Furthermore, we showed in \S \ref{Sdvdist} that in our sample the infall
signatures are seen almost entirely toward Class 0 rather than Class I sources.
The most straightforward interpretation of these results is that Class 0
sources differ from Class I sources in having a much greater incidence of
inward motions for the conditions which our observations probe --size scales of
a few $\times$ 0.01 pc and gas density of about 10$^5$ cm$^{-3}$.  This
difference, which has not been demonstrated before, adds support to the idea
that Class 0 sources have not yet accreted most of their mass, based mainly on
the greater submillimeter continuum emission of Class 0 than Class I sources
(Andre et al.\ 1993).  

The lack of infall signatures in Class I objects, however, does not necessarily
mean that infall is absent in these sources.  Adams, Lada, \& Shu (1987) showed
that to model the spectral energy distribution of Class I sources it is
necessary to include the contribution of an infalling envelope.  The lack of
blue-shifted line profiles toward Class I sources could therefore result from a
lower sensitivity of our observations to the physical conditions in the
infalling envelopes of Class I sources.  This would occur if the circumstellar
envelopes in Class I sources have less optical depth than in Class 0 sources,
or if the circumstellar envelopes in Class I sources have smaller excitation
temperature gradients than in Class 0 sources (in the optically thick CS and
H$_2$CO lines).

To test the above ideas, we examined the optical depths determined from the
multicomponent fits to the N$_2$H$^+$ spectra for the Class 0 and Class I
sources in our sample.  We found no significant difference between the two
groups: the N$_2$H$^+$ optical depth, summed over all seven hyperfine
components, has mean and s.e.m.\ $4.9 \pm 0.6$ for the Class 0, and $4.6 \pm
0.7$ for the Class I source sample.  This suggests that the greater incidence
of blue-shifted profiles in Class 0 than in Class I sources does not arise from
differences in line optical depth between the two samples.  However, the lack
of self-absorption features in the Class I source line profiles may indicate
that the CS and H$_2$CO optical depths could still be higher in Class 0 than in
Class I sources.    Also, there are more weak single-peaked sources in the
Class I than in the Class 0 source samples.

Sources whose outflow axes are in the plane of the sky, having edge on disks,
may have significant obscuration to optical and near-infrared radiation,
lowering their observed \tb.  This effect would tend to mix the Class 0 and
Class I sources.  However, the circumstellar envelopes are most likely
optically thin at sub-mm wavelengths, so a sub-mm wavelength excess should
distinguish Class 0 from Class I sources, independent of orientation.  Among
our sample of Class 0 sources, only 03282+3035 and L673A have not been detected
at sub-mm wavelengths; moving them to the Class I sample would only emphasize
the observed difference between the Class 0 and Class I samples.

Thus, we found a significant difference in the incidence of blue-shifted line
asymmetry between the Class 0 and Class I sources, indicating different
physical conditions in the circumstellar envelopes around Class 0 and Class I
sources.  However, our observations do not necessarily rule out the presence of
inward motions onto Class I sources; tracers with higher optical depths may be
needed to probe the circumstellar kinematics around Class I sources.

\subsection{Identification of Infall Candidates	\label{Scand}}

We summarize in table \ref{Tinfall} the evidence for infall, expansion, or
neither for the sources in our sample.  We list all 23 sources with $|\delta
V|>0.25$ in either the H$_2$CO or CS lines.  We label the lines b if they have
blue asymmetry ($\delta V < -0.25$), r if they have red asymmetry ($\delta V >
0.25$), or n if they have neither ($-0.25 \leq \delta V \leq 0.25$).  We
include IRAS 13036-7644 in the CS sample, using the spectrum published by
Lehtinen (1997).  The correspondance between both lines is extremely good, as
expected from figure \ref{Fh2cocs}.  Following Gregersen et al.\ (1997), we
identify infall candidates as those sources with evidence for infall (at least
one ``b'') and no evidence for expansion in any line (no ``r''s).  Therefore,
from the H$_2$CO and CS observations presented in this paper, we select 15
sources as infall candidates: IRAS 03256+3055, NGC1333-4A, NGC1333-4B, L1527,
IRAS 13036-7644, VLA1623, WL22, IRAS 16293-2422, L483, S68N, SMM5, SMM4, B335,
L1157, and L1251B.  These are Class 0 sources ($\tb < 70$~K) with the exception
of IRAS 03256+3055, WL22 and L1251B.  In contrast to the 15 (CS and H$_2$CO)
sources with at least one {\sl b} and no {\sl r}s, there are only 4 with the
opposite property --at least one {\sl r} and no {\sl b}s (NGC1333-2, L1551-5,
L1551NE and L43).

Among our infall candidates, L1527, IRAS 16293-2422 and B335 have been widely
studied (eg.\ Walker et al.\ 1986; Zhou et al.\ 1993, 1994; Mardones et al.\
1994; Myers et al.\ 1995; Ohashi et al.\ 1997; Gregersen et al.\ 1997).  They
represent the best cases for infall to date due to the agreement between
kinematic models and observations.  The sources NGC1333-2 (Ward-Thompson et
al.\ 1996), IRAS 13036-7644 (Lehtinen 1997), L483 (Myers et al.\ 1995), S68N
(Hurt, Barsony, \& Wootten 1996), SMM4 (Hurt et al.\ 1996, Gregersen et al.\
1997), L1157 (Gueth et al.\ 1997), and L1251B (Myers et al.\ 1996) have been
recently identified as infall candidates but have not yet been studied as
thoroughly as L1527, IRAS 16293-2422, and B335.  The sources IRAS 03256+3055,
NGC1333-4A and 4B, VLA1623, WL22, and SMM5 are new kinematic infall candidates
based on this work.

\subsection{Comparison with Previous Surveys}


Wang et al.\ (1995) observed 12 globules at the CSO using the H$_2$CO
$3_{12}-2_{11}$ line looking for kinematic infall signatures.  They do not find
convincing spectral evidence of infall toward any source.  This may be because
none of their sources are known to be Class 0 YSOs; or due to the expected
lower optical depth of the H$_2$CO $3_{12}-2_{11}$ line.

Gregersen et al.\ (1997) observed 23 Class 0 sources in the HCO$^+$ $3-2$ and
$4-3$ lines looking for infall signatures.  Their sample contains sources more
distant than 400~pc, but the overlap with our sample is still considerable,
with 15 sources in common.  We derived $\delta V$ from the HCO$^+$ and
H$^{13}$CO$^+$ spectra presented by Gregersen et al.\ toward 13 of those
sources, and assigned asymmetry labels b,r, and n to them in the same way as
for the CS and H$_2$CO spectra.  In table \ref{Tinfall} we list the asymmetry
labels from available observations: from this paper, from Gregersen et al.\
(1997), from Ward-Thompson et al.\ (1996), and from Lehtinen (1997).  The
correspondance among all lines is generally good.  Seven of the nine sources
with infall asymmetry in the HCO$^+$ lines have the same sense of asymmetry in
the H$_2$CO and CS lines.  On the other hand, the two sources with expansion
asymmetry in the HCO$^+$ lines show infall asymmetry in the H$_2$CO and/or CS
lines (L1448mm and L483), and one source (NGC1333-2) shows infall asymmetry in
HCO$^+$ $4-3$ (Ward-Thompson et al 1996) and expansion asymmetry in CS and
H$_2$CO.  These disagreements may reflect differing degrees of outflow
contamination from one line to the next.

Assuming as in \S \ref{Scand} that the best infall candidates are those sources
with evidence for infall in at least one line, and with no evidence for
expansion in any line, we consider again the sources in Table \ref{Tinfall},
now according to their asymmetry in HCO$^+$, as well as in CS, H$_2$CO.  Adding
this HCO$^+$ data and applying this procedure removes only one source, L483,
from the CS and H$_2$CO list given in \S \ref{Scand}, leaving the 14 sources
indicated in Table \ref{Tinfall} as the best infall candidates.  These sources
have substantial internal agreement among their asymmetry labels: 5 have 4 {\sl
b}s, 2 have 3 {\sl b}s, 5 have 2 {\sl b}s, and 2 have 1 {\sl b}.  At present,
these sources have the most evidence from available spectral line surveys for
infall, and they have the least evidence against infall.

\subsection{Thermal and Turbulent Infall}

The kinematic infall candidates identified in this paper include two sources
--B335 and L1527-- which exemplify in several ways the ``standard model'' of
low-mass star formation (e.g.\ Shu, Adams, \& Lizano 1987).  These sources are
relatively isolated and have NH$_3$ line widths dominated by thermal, rather
than turbulent motions (Myers, \& Benson 1983; Benson, \& Myers 1989).  Their
profiles of H$_2$CO and other lines are well-fit by radiative transfer models
whose kinematics are specified by the ``inside-out'' mode of collapse of a
singular isothermal sphere (Shu 1977).

But these two sources are not representative of the properties of the 15
kinematic infall candidates identified by the CS and H$_2$CO lines in this
paper.  Most of these sources differ from the standard model in two ways:
their line widths are dominated by turbulent motions, and many of them are
found in stellar groups rather than in isolation.

We quantify the turbulent motions of the infall candidates by comparing the
FWHM of the optically thin N$_2$H$^+$ line, given in Table \ref{Tlines}, with
that expected for gas having equal thermal and nonthermal motions,
\begin{equation}
\Delta V_0 = \sqrt{ 
8\ln{2} ~\mbox{kT} \left(\frac{1}{m_{obs}}+\frac{1}{\langle m \rangle}\right) 
             }
\end{equation}
where T is the gas kinetic temperature, $m_{obs}$ is the mass of the observed
species (29 amu for N$_2$H$^+$), and $\langle m \rangle$ is the mean molecular
mass (2.3 amu).  For gas with T = 15 K, 9 out of 15 sources, or 60\%, have
N$_2$H$^+$ line widths $\Delta V > \Delta V_0 = 0.57$ \kms, i.e. having
greater non-thermal than thermal motions.  If T = 10 K, this fraction increases
to 12/15, or 80\%.  Clearly most of the kinematic infall candidates have
non-thermal motions exceeding their thermal motions.

The non-thermal motions in the N$_2$H$^+$ line width could arise from motions
which appear random to our resolution, which we designate as turbulence, and
also from systematic motions, including outflow, infall, and rotation.  Since
N$_2$H$^+$ lines (like NH$_3$ lines) tend to trace quiescent core structure
(Womack et al.\ 1992) and do not track outflow maps (Caselli et al.\ 1997), we
expect that the outflow contribution to the N$_2$H$^+$ line width is
relatively small compared to the turbulent component.  Similarly, the relative
contribution of rotation and infall motions to the optically thin line width
is significantly smaller than the random contribution, according to infall
models which match the profiles of infall candidates (e.g.\ Zhou et al.\
1993).

To assess the proportion of the infall candidates associated with groups or
clusters, we consider those infall sources in the NGC1333, Ophiuchus, and
Serpens clouds, which are associated with a substantial cluster of embedded
young stellar objects.  Of the 15 infall candidates listed in Section \S
\ref{Scand}, 8 or 53\% belong to these three clusters.  This accounting is
conservative in that we do not count L1251B which belongs to a smaller group of
at least 5 near infrared sources (Hodapp 1994).  Of these 8 sources in
clusters, most (6) have N$_2$H$^+$ line widths dominated by turbulent motions.

This relatively high incidence of turbulent motions and cluster sources in our
list of kinematic infall candidates implies that the standard model of
isolated star formation in a thermally dominated core is not representative of
most of our infall candidates.  This is an area where more theoretical models
are needed.  Models of star formation which account for significant turbulent
motions as part of their initial conditions have been presented for a
spherically symmetric geometry and for the formation of a single star (Caselli
\& Myers 1995; McLaughlin \& Pudritz 1997); but these models lack a physical
basis for the non-thermal motions, and do not account for the formation of
multiple stars.

\subsection{Future Prospects \label{Sfut}}

The infall candidates identified in this paper are good candidates for the
study of star-forming infall.  However, it is not established from the survey
data presented here that their motions are necessarily either ``star-forming''
or ``gravitational.''

A further step to establish if the blue-shift identified in Class 0 sources
arise from infalling envelopes is to map them: the blue-shifted, asymmetrical
line profiles should be spatially concentrated toward the source.  Furthermore,
maps are needed to disentangle the effects of the outflows.  Maps of observed
line asymmetry (eg.\ Adelson \& Leung 1988; Walker et al.\ 1994) can
distinguish whether infall dominates (at low velocities) over rotation or
expansion in a given region.

To establish whether the observed infall kinematics are gravitational, the
observations should be compared with kinematic models, which predict the
spatial variation of the line profiles (eg.\ Zhou et al 1993).  Furthermore,
continuum observations may help determine the YSO and envelope masses to
compare with model predictions.  High spatial resolution attainable with
interferometers can locate the infall regions in relation to the position of
the continuum peak of the YSO.  All of these together should constrain
significantly kinematic models of gravitational collapse, to determine better
the physical basis of the inward motions inferred in this paper.  In addition,
further models and observations may be needed to resolve the apparent conflicts
presented in table \ref{Tinfall}.  Such models will have to incorporate both
infall and outflow kinematics.

\section{Conclusions}

Among 47 of the youngest nearby low mass PMS stars (spectral classes 0, or
$\tb < 70$~K, and I, or $70 \leq \tb \leq 200$~K) a significant fraction show
complex line profiles with double peaks or a peak and a shoulder in the
optically thick CS $2-1$ and H$_2$CO $2_{12}-1_{11}$ transitions, which trace
dense gas ($n >10^4$ cm$^{-3}$).

These complex spectral profiles are not due to multiple components along the
line of sight, since the optically thin N$_2$H$^+$ $101-012$ line, observed
toward the same sources, is single.  The N$_2$H$^+$ line velocities are
consistent with the C$^{34}$S 2-1 velocities.

In both CS or H$_2$CO samples, about half the sources have a significant
velocity-shift between the optically thick and thin lines.

In both CS or H$_2$CO samples, there are $\sim$25\% more sources whose velocity
shift is to the blue than to the red.  This statistically significant tendency
arises entirely from sources with $\tb < 70$~K (Class 0 sources).  In this
Class 0 sample the blue-shifted excess is 50\% (9 out of 23 for H$_2$CO and 10
out of 19 for CS), while in the Class I subsample there is no significant blue-
or red-shifted excess.

The observed excess of blue-shifts cannot be reproduced by a sample of sources
whose kinematics are dominated either by bipolar outflows or rotation, if their
symmetry axes are oriented in random directions.  On the other hand, infall
motions alone, or infall combined with rotation and/or outflows, can explain
the excess of blue-shifts.

We identify a total of 15 spectroscopic infall candidates, of which 6 have not
been identified previously.  Most of these candidates differ from the
well-known sources B335 and L1527, which exemplify the ``standard model'' of an
isolated core whose line width is dominated by thermal motions.  They tend to
have primarily turbulent line-widths, or are associated with embedded young
clusters, or both.

\acknowledgements
This research was supported by NASA Origins of Solar Systems Program, grant
NAGW-3401.  Support for this work was provided by NASA through Hubble
Fellowship grant HF-01086.01-96A awarded by STScI, which is operated by AURA,
Inc., for NASA under contract NAS 5-26555.  D.M.\ thanks the Carnegie
Institution of Washington for a Carnegie-Chile Fellowship and the Government of
Chile for a MIDEPLAN Fellowship.  M.T.\ and D.J.W. thank the
Harvard-Smithsonian Center for Astrophysics for fellowship support.  R.B.\ and
M.T.\ ackowledge partial support from the Spanish DGICYT grant PB93-48.


\clearpage

\clearpage

\begin{table}
\dummytable
\epsscale{1.16}
\plotone{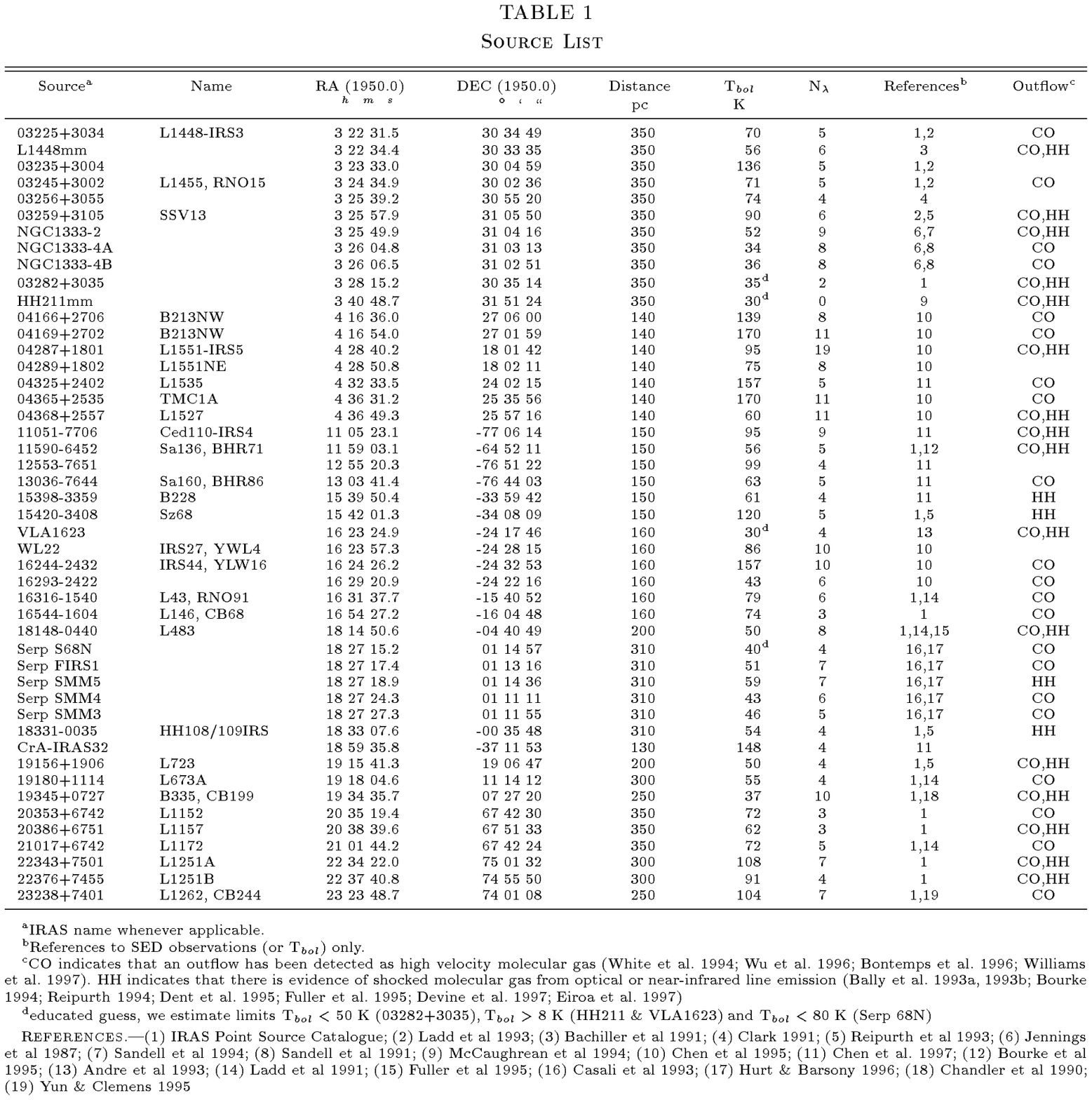}
\label{Tsources}
\end{table}

\begin{table}
\dummytable
\epsscale{1.16}
\plotone{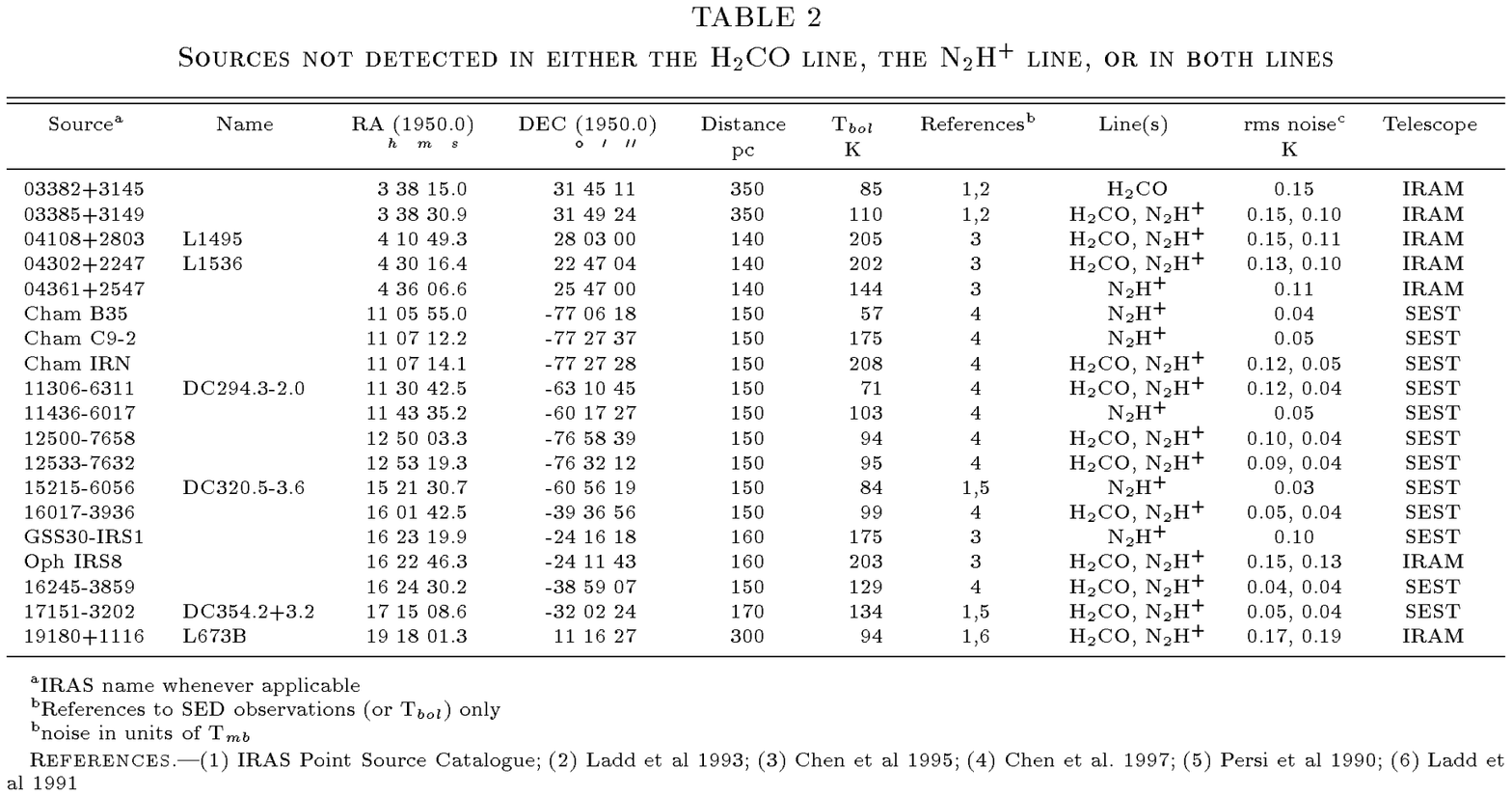}
\label{Tnodet}
\end{table}

\begin{table}
\dummytable
\epsscale{1.16}
\plotone{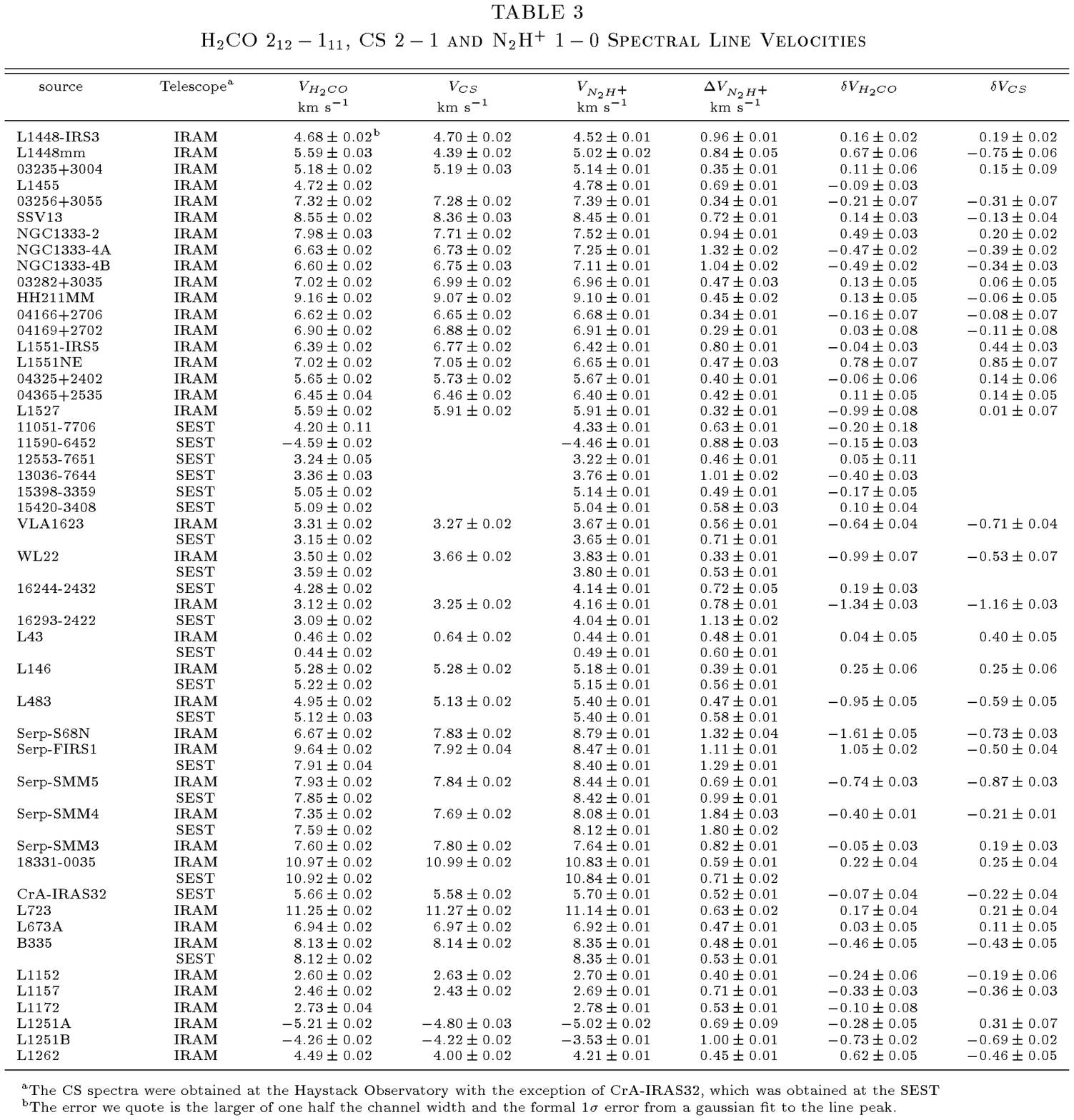}
\label{Tlines}
\end{table}

\begin{deluxetable}{lcrrc} 
\small
\tablecaption{H$_2$CO $2_{02}-1_{01}$ and C$^{34}$S $2-1$ Spectral Line Velocities 
	\label{Tisotope}}
\tablehead{
	\colhead{Source} & \colhead{Telescope} & 
	\colhead{$V_{H_{2}CO}$}	&
	\colhead{$V_{C^{34}S}$}	& \colhead{$\Delta V_{C^{34}S}$}	\nl
	\colhead{} & \colhead{} & 
	\colhead{km~s$^{-1}$} 	& \colhead{km~s$^{-1}$}   & \colhead{km~s$^{-1}$}
	}
\startdata
L1527      & IRAM &                 & $~5.94\pm 0.02$ & $0.30\pm 0.05$ \\
13036-7644 & SEST & $~3.42\pm 0.02$ & $~3.72\pm 0.04$ & $0.53\pm 0.10$ \\
VLA1623    & SEST & $~3.54\pm 0.02$ & $~3.58\pm 0.05$ & $1.37\pm 0.12$ \\
WL22       & SEST & $~3.58\pm 0.02$ & $~3.87\pm 0.02$ & $0.74\pm 0.04$ \\
16293-2422 & SEST & $~3.22\pm 0.02$ & $~4.02\pm 0.06$ & $1.78\pm 0.21$ \\
L483       & SEST & $~5.31\pm 0.05$ & $~5.40\pm 0.02$ & $0.71\pm 0.05$ \\
Serp-FIRS1 & SEST & $~7.96\pm 0.02$ & $~8.53\pm 0.03$ & $1.32\pm 0.08$ \\
Serp-SMM5  & SEST & $~7.87\pm 0.07$ & $~8.40\pm 0.03$ & $0.97\pm 0.09$ \\
Serp-SMM4  & SEST & $~7.55\pm 0.02$ & $~8.11\pm 0.05$ & $2.07\pm 0.21$ \\
18331-0035 & SEST & $10.82\pm 0.02$ & $10.86\pm 0.03$ & $0.37\pm 0.07$ \\
CrA IRS32  & SEST & $~5.69\pm 0.02$ & $~5.72\pm 0.05$ & $0.72\pm 0.11$ \\
B335       & SEST & $~8.16\pm 0.02$ & $~8.43\pm 0.04$ & $0.59\pm 0.10$ \\
L1251B     & IRAM &                 & $-3.74\pm 0.05$ & $1.42\pm 0.12$ \\
\enddata
\end{deluxetable}

\clearpage

\begin{deluxetable}{clccccccc} 
\tablecaption{Statistical properties of the distributions of
	$\delta V$ \label{Tstat}}
\footnotesize
\tablehead{
	\colhead{Optically thick line}	& \colhead{Sample} & \colhead{N} & 
	\colhead{N$_{-}$\tablenotemark{a}} & \colhead{N$_{0}$}	& \colhead{N$_{+}$}	&
	\colhead{mean $\pm$ s.e.m.} & 	
	\colhead{p\tablenotemark{b}} & \colhead{E\tablenotemark{c}}	
	}
\startdata
\tablevspace{0.5mm}
H$_2$CO	& all                       & 47 & 15 & 27 & 5 & $ -0.14\pm 0.08$ & 0.06\phn & ~0.21 \\
	\tablevspace{0.8mm}		       	      
	& $T_{bol}\leq 70$~K        & 23 & 12 & ~8 & 3 & $ -0.28\pm 0.13$ & 0.05\phn & ~0.39 \\
	& $70<T_{bol}<200$~K        & 24 & ~3 & 19 & 2 & $ -0.02\pm 0.07$ & 0.81\phn & ~0.04 \\
					       	      
\nl					       	      
CS	& all                       & 37 & 14 & 19 & 4 & $ -0.14\pm 0.07$ & 0.04\phn & ~0.27 \\
	\tablevspace{0.8mm}		       	      
	& $T_{bol}\leq 70$~K        & 19 & 10 & ~9 & 0 & $ -0.28\pm 0.10$ & 0.008    & ~0.53 \\
	& $70<T_{bol}<200$~K        & 18 & ~4 & 10 & 4 & $~~0.00\pm 0.09$ & 0.99\phn & ~0.00 \\
\enddata

\tablenotetext{a}{N$_{-}$, N$_{0}$, N$_{+}$ are the number of sources in the subsample with 
     	normalized velocity difference $\delta V<-0.25$, $-0.25 \leq \delta V \leq 0.25$, 
	and $0.25 < \delta V$ respectively}
\tablenotetext{b}{Probability of drawing the sample from a zero-mean normal parent distribution,
	based on a Student's t-distribution}
\tablenotetext{c}{The blue excess is defined as $E = (N_{-}-N_{+})/N$}
\end{deluxetable}

\clearpage

\begin{center}
\begin{deluxetable}{lccccc} 
\tablecaption{Line Asymmetries in Sources Having $|\delta V|>0.25$ in at Least
	One Line	\label{Tinfall}}
\footnotesize
\tablehead{
        \colhead{Source\tablenotemark{a}}  & \colhead{T$_{bol}$} &
	\multicolumn{4}{c}{Line Asymmetry\tablenotemark{b}} \nl 
	\cline{3-6} \tablevspace{0.8mm}
        \colhead{} & \colhead{K} & 
	\colhead{CS $2-1$} & \colhead{H$_2$CO $2_{12}-1_{11}$} & 
	\colhead{HCO$^+$ $3-2$} & \colhead{HCO$^+$ $4-3$} 
	  }
\startdata
L1448mm          & \phm{1}56 & b & r & r & r \nl
{\bf 03256+3055} & \phm{1}74 & b & n &   &   \nl
N1333-2          & \phm{1}52 & r & r &   & b\tablenotemark{(1)} \nl
{\bf N1333-4A}   & \phm{1}34 & b & b & b & b \nl
{\bf N1333-4B}   & \phm{1}36 & b & b & b & b \nl
L1551-5          & \phm{1}95 & n & r &   &   \nl
L1551NE          & \phm{1}75 & r & r & b &   \nl
{\bf L1527   }   & \phm{1}60 & n & b & b & b \nl
{\bf 13036-7644} & \phm{1}63 & b\tablenotemark{(2)} & b       \nl
{\bf VLA1623 }   & \phm{1}30 & b & b & n & n \nl
{\bf WL22    }   & \phm{1}86 & b & b &   &   \nl
{\bf 16293-2422} & \phm{1}43 & b & b & b & b \nl
L43	         & \phm{1}79 & n & r &   &   \nl
L483             & \phm{1}50 & b & b & r & r \nl
{\bf S68N    }   & \phm{1}40 & n & b &   &   \nl
FIRS1            & \phm{1}51 & b & r & r & b \nl
{\bf SMM5    }   & \phm{1}59 & b & b &   &   \nl
{\bf SMM4    }   & \phm{1}43 & b & b & b & b \nl
{\bf B335    }   & \phm{1}37 & b & b & b & b \nl
{\bf L1157   }   & \phm{1}62 & b & b & b & n \nl
L1251A	         & 108       & b & r &   &   \nl
{\bf L1251B  }   & \phm{1}91 & b & b &   &   \nl
L1262            & 104       & r & b &   &   \nl
\enddata			  
				  
\tablenotetext{a}{The best cases for infall are indicated in bold face}
\tablenotetext{b}{The line asymmetry is designated b for $\delta V< -0.25$, n
for $-0.25\leq \delta V \leq 0.25$, and r for $\delta V > 0.25$.    Except where
noted, the CS and H$_2$CO data come from this paper, and the HCO$^+$ line data
come from Gregersen et al.\ (1997)}
\tablerefs{(1) Ward-Thompson et al 1996, (2) Lehtinen 1997}
\end{deluxetable}
\end{center}


\end{document}